\documentclass[letterpaper,aps,10pt,prb,twocolumn,showpacs,superscriptaddress]{revtex4-1}

\usepackage{graphicx}
\usepackage{hyperref}
\usepackage{amsmath}
\usepackage{amssymb}
\usepackage{txfonts}
\makeatletter

\begin{document}
\newcommand{\sign}{\mathrm{sign}\,}
\newcommand{\figref}[1]{Fig.~\ref{#1}}
\newcommand{\expct}[2]{\left\langle #1 \right\rangle_{#2}}
\newcommand{\expcts}[2]{\langle #1 \rangle_{#2}}
\newcommand{\cumu}[2]{\langle\!\langle #1 \rangle\!\rangle_{#2}}
\newcommand{\C}{\mathcal{C}}
\newcommand{\T}{\mathcal{T}}
\newcommand{\hQ}{\lw{replace}}
\newcommand{\Tr}{\mathrm{Tr}}
\newcommand{\TC}{T_{\C}\, }
\newcommand{\hc}{\text{h.c.}}
\renewcommand{\vec}[1]{{#1}}
\newcommand{\mum}{\mu\mathrm{m}}
\newcommand{\mueV}{\mu\mathrm{eV}}
\newcommand{\mK}{\mathrm{mK}}
\newcommand{\eff}{\mathrm{eff}}
\newcommand{\eqrefmanual}[1]{Eq.~(#1)}

\title{Electron transport in multiterminal networks of Majorana bound states}
\author{Luzie Weithofer}
\affiliation{Institute for Mathematical Physics, TU Braunschweig, 38106 Braunschweig, Germany}

\author{Patrik Recher}
\affiliation{Institute for Mathematical Physics, TU Braunschweig, 38106 Braunschweig, Germany}

\author{Thomas L. Schmidt}
\affiliation{Department of Physics, University of Basel, Klingelbergstrasse 82, 4056 Basel, Switzerland}

\date{\today}

\begin{abstract}
We investigate electron transport through multiterminal networks hosting Majorana bound states (MBS) in the framework of full counting statistics
(FCS). In particular, we apply our general results to T-shaped junctions of two Majorana nanowires. When the wires are in the topologically nontrivial regime, three MBS are localized near the outer ends of the wires, while one MBS is localized near the crossing point, and
when the lengths of the wires are finite adjacent MBS can overlap. We propose a combination of current and cross-correlation
measurements to reveal the predicted coupling of four Majoranas in
a topological T~junction. Interestingly, we show that the elementary transport
processes at the central lead are different compared to the outer leads,
giving rise to characteristic non-local signatures in electronic transport. We find quantitative agreement between our analytical model and numerical simulations of a tight-binding model. Using the numerical simulations, we discuss the effect of weak disorder on the current and the cross-correlation functions.
\end{abstract}
\maketitle

\section{Introduction}

The investigation of Majorana fermions has recently evolved into a vibrant
field within condensed-matter physics.
\cite{alicea12,beenakker13} Semiconductor nanowires with strong spin-orbit
coupling in proximity to a superconductor and subject to a magnetic field have been proposed as a platform for the realization of Majorana bound states (MBS).\cite{Oreg2010,Sau2010} The low-energy sector of these systems is identical to that of a one-dimensional spinless $p$-wave superconductor, which can be described as a Kitaev chain.\cite{Kitaev2001} Recent experiments\cite{Mourik2012,Deng2012,Das2012,Lee2014} indeed provide first possible signatures of MBS.

It was theoretically predicted that MBS can be an important resource for topological quantum
computation, which promises to facilitate qubit operations that are protected from certain common sources of decoherence.\cite{Nayak2007} The key qubit operation is braiding of MBS, which can be achieved in T~junctions of one-dimensional Majorana wires.\cite{Alicea2010} Such junctions should be experimentally realizable using crossed wires, and they are fascinating in their own
right. For instance, since MBS always occur in pairs, a MBS must also emerge at the crossing point
of a topological T~junction, and not only at the ends of
the nanowires. This is indeed what an extension of the Kitaev model
to two crossed chains predicts\cite{Alicea2010} (see also Appendix~\ref{appendix:Kitaev}).

In this paper, we present the spectral properties and the transport
  properties of a topological
T~junction that result from the hybridization of adjacent MBS when the wires
are of finite length. As we will show, the electronic transport signatures of
the central MBS in such a T~junction are very different from those of the outer
MBS. In particular, we show that a measurement of current cross-correlations involving the central MBS reveals clearly the nonlocality of transport.
This characteristic effect
  could serve as a hallmark for understanding topological T~junctions.
Our theory is particularly timely
in the light of recent experimental progress with crossed Majorana nanowires,\cite{vanweperen13} and recent theoretical proposals for HgTe quantum wells
as topological superconductors\cite{Weithofer13,Reuther2013} and their
patterning into wires.\cite{Reuther2013}

Using full counting statistics (FCS), we investigate transport
in general networks of pairwise coupled MBS, where each MBS can be connected to a metallic lead by
tunneling. FCS yields not only the average current at each contact, but also arbitrary
higher-order current correlation functions. Moreover, it allows us to identify the elementary transport processes in all parameter regimes. First, we use our general result~\eqref{eq:generalFCS} to physically interpret the known results\cite{Nilsson2008,wu12b,Li2012,bin-he2014} for
a Majorana nanowire with two MBS at the ends. Subsequently, we discuss in detail the transport properties of a T~junction of Majorana
nanowires. We find that at zero bias voltage, the elementary transport
processes have no trivial equivalent in the previously described setup of a Majorana
nanowire with two MBS at the ends. They are a combination of resonant concurring crossed Andreev reflection (CAR) processes at the
outer leads and a non-resonant process involving the central lead which we call double CAR. We show that this has a strong impact on
current cross-correlations, and can be used to probe the nonlocal nature of
transport through MBS. We show that these cross-correlations provide a
stronger signature of MBS-based transport than conductance measurements. In addition, we present numerical results for a T~junction realized in a weakly disordered semiconductor nanowire, and find quantitative agreement with our analytical predictions.

\section{FCS in a network of MBS.}
We consider a general network of $N$ localized MBS $\gamma_\alpha$. As the lengths of the wire segments between MBS are
finite, a nonzero overlap of the MBS $\gamma_\alpha$ and $\gamma_\beta$ causes an energy splitting $\varepsilon_{\alpha\beta}$. In addition, we assume that each MBS can be connected to a normal-metal lead via electron tunneling. Each MBS couples to a linear combination of spin-up and spin-down electrons in the leads, whereas the orthogonal linear combination remains uncoupled.\cite{law09} Therefore, we can model the system by the spinless Hamiltonian $H = H_{\rm leads} + H_M + H_{\rm tun}$, where (using $e = \hbar = 1$)
\begin{align}
   H_{\rm leads} &= -i v_F  \sum_{\alpha=1}^N  \int dx \psi_\alpha^\dag(x) \partial_x \psi_\alpha(x), \notag \\
H_M &= -\frac{i}{2} \sum_{\alpha\neq\beta} \varepsilon_{\alpha\beta}
    \gamma_\alpha \gamma_\beta, \notag \\
    H_{\rm tun} &= -i \sum_{\alpha=1}^N \gamma_\alpha \left[ t_\alpha \psi_\alpha^\dag(0) + t_\alpha^* \psi_\alpha(0) \right]. \notag
\end{align}
Here, $\varepsilon_{\beta\alpha} = -\varepsilon_{\alpha\beta}$, $t_\alpha$ is the tunneling amplitude between the MBS $\gamma_\alpha$ and the lead $\alpha$, and $\psi_\alpha^\dagger(x)$ denotes the electron creation operator in that lead.

The goal of FCS is the calculation of the
cumulant generating function (CGF) $\ln \chi$, which can be represented as a
time-ordered expectation value on the Keldysh contour,\cite{Levitov2004}
\begin{align}\label{eq:lnchi_K}
 \ln \chi(\lambda) &= \ln \expct{ \TC \exp \left[ - i \int_\C ds
     H^{\lambda}(s) \right] }{} ,\\
H^\lambda(s)&=H +
\sum_\alpha\frac{\lambda_{\alpha}(s)}{2}\,\hat{I}_\alpha.\notag
\end{align}
where $\expct{\dots}{}$ denotes the quantum mechanical average with
respect to the equilibrium density matrices of the leads and the Majorana system, respectively. 
Here, $\TC$ denotes the time-ordering operator\cite{lifschitz81} on the Keldysh contour $\C$,
which consists of two branches: $\C_-$ leads from $-\infty$ to $+\infty$
whereas $\C_+$ goes back to $-\infty$.

In principle, the CGF is the generating function for the current operator $\hat{I}_\alpha =
d\hat{N}_\alpha/dt$ at lead $\alpha$, e.\,g. its expectation value
$\expcts{\hat{I}_\alpha(t)}{}$ at a certain time $t$ can be
obtained by taking the variational derivative of the CGF by the ``counting
field'' $\lambda_\alpha(t)$. Here, the number operator in lead $\alpha$ is
given by $\hat{N}_\alpha=\int dx\,\psi_\alpha^\dagger(x)\psi_\alpha(x)$.
As we are interested in counting statistics of the charge transmitted over a long measurement time
$\T$, we define the counting fields as
$\lambda_{\alpha}(s) = \pm\lambda_{\alpha}\theta(s)\theta(\T-s)$ for $s \in
\C_{\pm}$\cite{Levitov2004}, where $\theta(x)$ is the Heaviside
function. Then, derivatives of the generating function $\chi(\lambda)$ with
respect to $\lambda_\alpha$ generate cumulants of charge, so that we can write
equivalently to Eq.~\eqref{eq:lnchi_K}
\begin{align}
    \chi(\lambda)= \sum_{\{q_\alpha\}} e^{-i\sum_\alpha q_\alpha\lambda_\alpha}\mathcal{P}(\{q_\alpha\}),
\end{align}
where $\mathcal{P}(\{q_\alpha\})$ is the joint probability of a charges $q_\alpha$ being
transferred at the respective leads $\alpha$ within the measurement time $\T$.
The transport properties
such as the average currents $I_\alpha$ and the zero-frequency limit of the corresponding
symmetrized current-current correlation function $P_{\alpha\beta}$ are easily deduced from
the CGF by
\begin{align}
  I_\alpha &= \frac{i}{\T} \frac{\partial \ln \chi}{\partial \lambda_\alpha} \bigg|_{\vec{\lambda}=0}= \frac{1}{\T} \int_0^\T dt\,\expcts{\hat{I}_\alpha(t)}{} ,\label{eq:currentcalc}\\
P_{\alpha\beta}&= - \frac{1}{\T}\frac{\partial^2 \ln \chi}{\partial
  \lambda_\alpha\partial \lambda_\beta}\bigg|_{\vec{\lambda}=0} \approx  \frac{1}{2}\int_{-\T/2}^{\T/2} dt\,\expcts{\{\delta\hat{I}_\alpha(t),\delta\hat{I}_\beta(0)\}}{}
. \notag
\end{align}
where $\delta\hat{I}_\alpha(t)=\hat{I}_\alpha(t)-I_\alpha$ and the second
equation is a good approximation for long measurement times $\T$. 
These quantities lead
directly to the Fano factor, $F_{\alpha\beta}=2P_{\alpha\beta}/(I_\alpha+I_\beta)$, which is measurable in experiments and can further elucidate the nature of
the transport processes. Hence, FCS provides access to all zero-frequency transport properties of the system.
The CGF can be
calculated by following an approach demonstrated in Refs.~\onlinecite{Nazarov2003,Levitov2004,Gogolin2006,schmidt07,schmidt11}, which is based on an extension of the conventional Keldysh
technique. Using this procedure, one finds that the CGF can be
expressed in terms of the Keldysh Majorana Green's function matrix
$D^\lambda(\omega)$ (see Appendix~\ref{appendix:FCS_calculation}),
\begin{equation}
\ln \chi(\lambda)
=\frac{\T}{2}\int\frac{d\omega}{2\pi}\ln\frac{\det [D^{\vec{\lambda}}(\omega)]^{-1}}{\det [D^{\vec{\lambda}}(\omega)]^{-1}\big|_{\vec{\lambda}=0}}.
\label{eq:generalFCS}
\end{equation}
The Majorana Green's function can be determined
via the Dyson equation $[D^{\vec{\lambda}}]^{-1}
=[D^0]^{-1}-\Sigma^{\vec{\lambda}}$ from the unperturbed Majorana Green's function $D^0$
and the tunneling self-energy $\Sigma^{\vec{\lambda}}$, as described in detail
in the Appendix~\ref{appendix:FCS_calculation}.
Equation~\eqref{eq:generalFCS} is reminiscent of the generalized version of the Levitov-Lesovik
determinant formula\cite{Levitov1993} for fermions and multiple normal
leads and for a single normal lead has the form of a normal-superconducting contact\cite{Muzykanskii1994} in the
subgap regime. Similar expressions were found in the context of tunneling from a
single lead to a chain of MBS,\cite{Golub2011} albeit in the limit of small
counting fields.\par
For zero temperature and for symmetric bias $\mu_\alpha=\mu$ in the normal leads, one can show that this result can be simplified to
\begin{align}
    \ln \chi(\lambda)=\T\int_{-\mu}^\mu \frac{d\omega}{2\pi}\ln
    \det[1-G_R(\omega)V_R(\omega)],\label{eq:FCSzeroTemp}
\end{align}
where the determinant is to be understood only over lead indices $\alpha,\beta$, and
\begin{align}
[G_R(\omega)^{-1}]_{\alpha\beta}&=\delta_{\alpha\beta}\left(\frac{\omega}{2}+i\Gamma_\alpha\right)+i\epsilon_{\alpha\beta},\\
[V_R(\omega)]_{\alpha\beta}&=i\delta_{\alpha\beta}\Gamma_\alpha (e^{-i\lambda_\alpha}
-1),
\end{align}
where $\Gamma_\alpha=|t_\alpha|^2/ v_F $ is the tunneling rate at lead $\alpha$. From this formula, all transport properties can be easily calculated.

\section{Majorana wire with two leads.}
First, we apply FCS to a setup with two MBS connected to two normal-metal
leads. To obtain compact results, we assume that both leads have zero
temperature and are biased at identical chemical potentials $\mu$. Electron transport via the MBS occurs between the leads and the grounded superconductor.\cite{Bolech2007} The coupling between the MBS is given by $H_M=-i\varepsilon_{12} \gamma_1\gamma_2$. The transport properties of such a system have been described before,\cite{Nilsson2008,wu12b,Li2012,bin-he2014} but not in the framework of FCS. From Eq.~\eqref{eq:FCSzeroTemp}, we find the CGF
\begin{align}
&\ln \chi(\lambda)=\T \int_{-\mu}^\mu \frac{d\omega}{2\pi}\ln \left[\varepsilon_{12}^2-\left(\frac{\omega}{2}+i\Gamma_1 e^{-i\lambda_1}\right) \left(\frac{\omega}{2}+i\Gamma_2 e^{-i\lambda_2}\right)\right].\notag
\end{align}
In order to interpret the CGF physically, we approximate it (i)~for the case of zero bias ($\mu = 0$), (ii) on resonance for weak tunneling ($\Gamma_{1,2} \ll \mu = \pm 2 \varepsilon_{12}$), and (iii) in the limit of large bias $\mu \gg \varepsilon_{12}$. We obtain, in the respective limits, for the differential CGF
\begin{align}
\frac{\partial}{\partial \mu}\ln \chi\big|_{\mu= 0}&=\frac{\T}{\pi}
\ln\left[1+p(e^{-i(\lambda_1+\lambda_2)}-1)\right],\label{eq:CAR}\\
\frac{\partial}{\partial \mu}\ln
\chi\big|_{\mu^2\approx4\varepsilon^2_{12}}&\approx \frac{\T}{\pi} \ln
    \left[q e^{-i\lambda_2} +(1-q)e^{-i\lambda_1}\right], \label{eq:QR}\\
\frac{\partial}{\partial \mu} \ln \chi\big|_{\mu\gg\varepsilon_{12}}&\approx
\frac{\T}{2\pi} \sum_{\alpha=1}^2\ln
    \left[1+r_\alpha(e^{-2i\lambda_\alpha}-1)\right], \label{eq:LMMAR}
\end{align}
where the probabilities of the elementary transport processes are given by $p=\Gamma_1\Gamma_2/(\varepsilon_{12}^2+\Gamma_1\Gamma_2)$, $q=\Gamma_2/(\Gamma_1+\Gamma_2)$ and $r_\alpha=\Gamma_\alpha^2/[(\mu/2)^2+\Gamma_\alpha^2]$.

Equation~\eqref{eq:CAR} is the CGF of a binomial distribution and indicates that the elementary transport
process at $\mu = 0$ consists in transferring two fermions, one from lead 1 and one
from lead 2, into the superconductor with probability $p$, thereby
forming a Cooper pair. This process is called crossed Andreev reflection (CAR) and
was previously reported in Ref.~[\onlinecite{Nilsson2008}].

The differential CGF in Eq.~\eqref{eq:QR} indicates that the elementary charge transport process at resonance consists
of transferring a single fermion either from lead 2 (with probability
$q$) or from lead 1 (with
probability $1-q$) onto the fermionic resonant level (RL) formed by the pair
of MBS. A second fermion arrives at the RL when this process is repeated
independently, and a Cooper pair can form. This process has previously
  been described in slightly different setups as leading to
  antibunching\cite{Strübi2011} and to a ``Hanbury-Brown-Twiss effect'' in pseudospin space.\cite{Bose2011}

Lastly, the elementary processes described by Eq.~\eqref{eq:LMMAR}, which represent two independent binomial distributions, consist of
transferring two particles from a single lead $\alpha$ to the superconductor
with probability $r_\alpha$, thereby forming a Cooper pair. This is referred to as local Andreev
reflection (LAR). In this regime, the tunnel processes on opposite ends of the wire are statistically
independent, and have previously been
described in Ref.~[\onlinecite{Bolech2007}]. For $\varepsilon_{12} = 0$ and $\mu \to 0$, one finds $r_\alpha = 1$ and Eq.~(\ref{eq:LMMAR}) yields the quantized zero-bias peak in the differential conductance, $\partial_\mu I_\alpha\big|_{\mu = 0} = 1/\pi = 2e^2/h$.

\begin{figure}[t]
    \includegraphics[width=0.7\columnwidth]{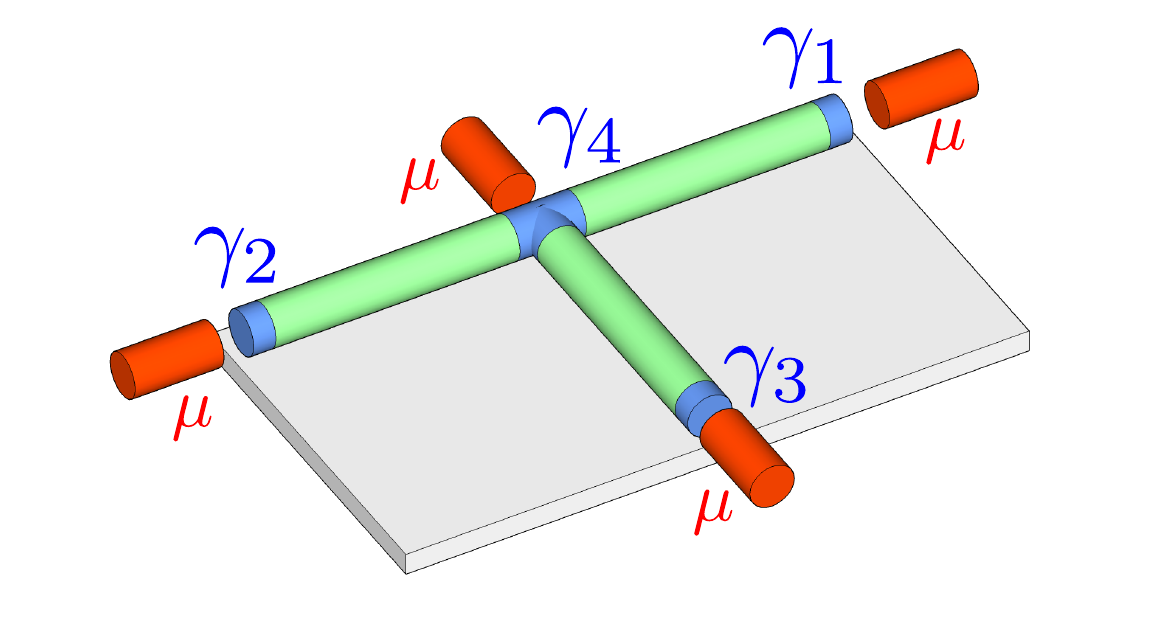}
    \centering
    \caption{Four Majorana bound states $\gamma_{1,2,3,4}$ (MBS, blue) in a T~junction are located at the ends of the superconducting wires (green), as well as at the crossing point. Each of the MBS is tunnel coupled to a normal-metal lead (red) with chemical potential $\mu$. The superconductor (grey) is grounded.}
    \label{fig:fourleads}
\end{figure}

\section{T~junctions}
In the following, we use the general result~\eqref{eq:FCSzeroTemp} to discuss the transport properties of a
T~junction consisting of two Majorana nanowires, see \figref{fig:fourleads}. This setup could be experimentally realized, e.g., by patterning HgTe quantum wells\cite{Reuther2013} in the shape of a T~junction or by using crossed semiconductor nanowires.\cite{vanweperen13} The low-energy sectors of these systems are identical to that of two Kitaev chains with a phase difference of $\pi/2$,\cite{Alicea2010} where the sites at the crossing point are identified. The ground state of the system is characterized by four Majoranas: one Majorana ($\gamma_4$) lies in close proximity to the crossing of the wires, while three other Majoranas ($\gamma_{1,2,3}$) are located at the ends of the
three arms\cite{Alicea2010} (see Appendix~\ref{appendix:Kitaev}). Each of
the outer Majoranas has a nonzero overlap with the central one, as described by the effective Hamiltonian
\begin{equation}
H_M=-i(\varepsilon_{14}\gamma_1\gamma_4+\varepsilon_{24}\gamma_2\gamma_4+\varepsilon_{34}\gamma_3\gamma_4).
\label{eq:effective4}
\end{equation}
We have neglected next-to-nearest-neighbor Majorana couplings because they are
exponentially suppressed compared to nearest-neighbor couplings. In addition,
each of the MBS is coupled to a normal-metal lead,
see~\figref{fig:fourleads}. \par
For the ensuing discussion, we assume equal bias voltages ($\mu_\alpha = \mu$) and zero temperature. As in Eqs.~\eqref{eq:CAR}-\eqref{eq:LMMAR}, we distinguish transport for zero bias, at resonance, and for large bias. In the limit of large bias voltage, i.e., $\mu \gg \varepsilon_{\alpha\beta}$, we obtain a straightforward generalization of Eq.~\eqref{eq:LMMAR} for the CGF. It consists of statistically independent binomial distributions for the four contacts, and LAR is the dominant transport mechanism at each contact.

\begin{figure}[t]
    \includegraphics[width=\columnwidth]{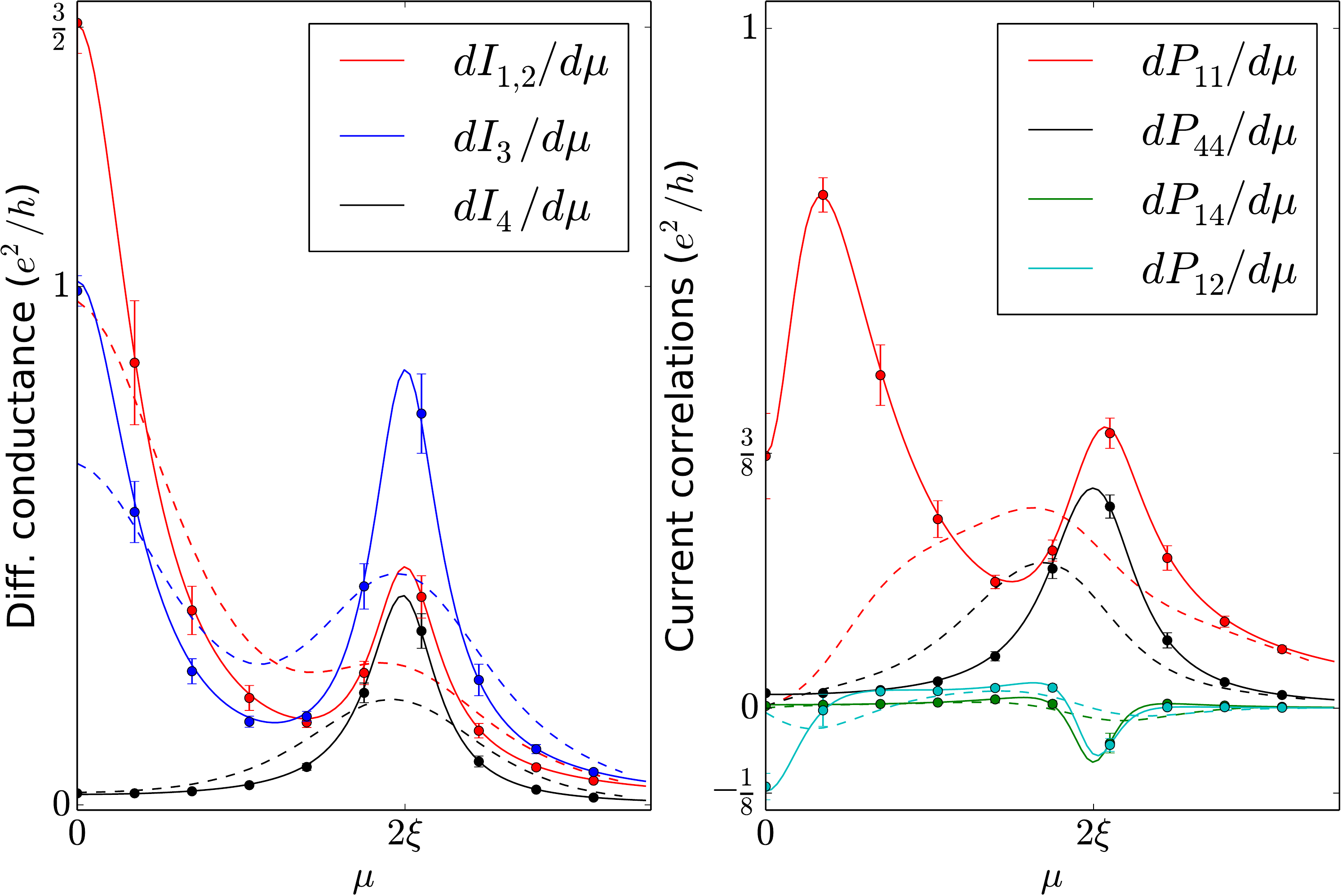}
    \centering
    \caption{Differential conductance $dI_\alpha/d\mu$ and current correlation
      functions $dP_{\alpha\beta}/d\mu$ for a
      T~junction with four leads. The solid (dashed)
      lines indicate calculations based on FCS determinant
      formula Eq.~\eqref{eq:generalFCS} applied to the low-energy
      Hamiltonian Eq.~\eqref{eq:effective4} with parameters $\varepsilon_{14} =
      \varepsilon_{24} = \xi/2, \varepsilon_{34} = \xi/\sqrt{2}$,
      $\Gamma_{1,2,3} = \Gamma, \Gamma_4 \approx \Gamma/4, \Gamma\approx
      \xi/5$ at zero temperature (temperature $kT=\xi/8$). For comparison,
      numerical results for a realistic T~junction setup with weak onsite
      disorder are also shown (dots with error bars), see Appendix~\ref{appendix:MicroscopicModel} for details.}
    \label{fig:fourleadscond}
\end{figure}

Resonant transport in the T~junction occurs at bias voltage $\mu \approx \pm 2 \xi$, where $\xi = (\varepsilon_{14}^2+\varepsilon_{24}^2+\varepsilon_{34}^2)^{1/2}$. In this RL regime, the differential CGF reads (for weak tunneling $\Gamma_\alpha \ll \epsilon_{\alpha\beta}$)
\begin{align}
    \frac{\partial}{\partial \mu}\ln \chi\big|_{\mu^2\approx 4\xi^2}&\approx \frac{\T}{\pi} \ln
    \left(\sum_{\alpha=1}^4 p_\alpha e^{-i\lambda_\alpha}\right),\label{eq:QR2}
\end{align}
where
\begin{align}
&p_{\alpha}=\frac{\Gamma_\alpha\varepsilon_{\alpha4}^2}{\Gamma_1\varepsilon_{14}^2+\Gamma_2\varepsilon_{24}^2+\Gamma_3 \varepsilon_{34}^2 + \Gamma_4\xi^2},& (\alpha = 1,2,3) \notag \\
&p_4=\frac{\Gamma_4\xi^2}{\Gamma_1\varepsilon_{14}^2+\Gamma_2\varepsilon_{24}^2+\Gamma_3 \varepsilon_{34}^2 + \Gamma_4\xi^2} .
\end{align}
Similarly to Eq.~\eqref{eq:QR}, this CGF indicates that an elementary charge process
consists of the transport of a single fermion onto the RL from one of the four
leads with respective probabilities $p_\alpha$, and when a second
fermion arrives independently, they can form a Cooper pair. Although the differential conductance through a single lead $\alpha$,
$\partial_\mu I_\alpha \approx 2e^2 p_\alpha/h$ is not quantized in this parameter regime, the sum of
the differential conductances of the single leads is quantized
\begin{equation}\label{eq:dIdmu_sum}
\frac{\partial (I_1+I_2+I_3+I_4)}{\partial \mu}\bigg|_{\mu^2\approx 4\xi^2}\approx\frac{2e^2}{h}.
\end{equation}
The quantization becomes exact for $\Gamma_\alpha \to 0$. Note that the
probability of an electron to arrive differs considerably at the central MBS
$\gamma_4$ compared to the outer MBS $\gamma_{1,2,3}$. The individual currents $I_\alpha$ fluctuate, but in a correlated way, because
a fermion leaving through lead $\alpha$ implies that no fermion has left
through any other lead. This is reflected in the maximally negative differential cross-correlations $\partial_\mu P_{\alpha\beta}
|_{\mu^2\approx 4\xi^2} =\frac{2e^2}{h} (p_\alpha \delta_{\alpha\beta}-
p_\alpha p_\beta)$. A plot
of these differential cross-correlations is shown in
\figref{fig:fourleadscond}.

Finally, we consider the CGF in the CAR limit $\mu \to 0$ at finite $\epsilon_{\alpha\beta}$. We find
\begin{align}
\frac{\partial}{\partial \mu} \ln \chi\big|_{\mu= 0} &=
\frac{\T}{\pi} \ln \left[ \frac{z_{12} + z_{23} + z_{31} + z_4}{z_0} \right], \label{eq:MR}
\end{align}
where $z_0 = (z_{12} + z_{23} + z_{31} + z_4)|_{\lambda = 0}$, $z_{12} = \Gamma_1 \Gamma_2 \epsilon_{34}^2 e^{-i(\lambda_1+\lambda_2)}$, and
$z_{23}$, $z_{31}$ are obtained by cyclic permutations of the indices $1,2,3$ in the
formula for $z_{12}$. We interpret the term $z_4 =\Gamma_1\Gamma_2\Gamma_3\Gamma_4e^{-i(\lambda_1+\lambda_2+\lambda_3+\lambda_4)}$ in the CGF (\ref{eq:MR}) as describing a \emph{double} crossed Andreev reflection process at the central
  lead which involves transport at each of the four leads, while the terms $z_{12},z_{23},z_{31}$ in the CGF~(\ref{eq:MR}) can be interpreted as describing
resonant concurring CAR processes involving only the three outer
normal leads: For example, an incoming electron in lead 3 that is not
backreflected, can be either emitted as a hole at lead 1 with probability
$z_{31}/z_0$ or be emitted as a hole at lead 2 with probability $z_{23}/z_0$,
whereby a Cooper pair is deposited in the superconductor. Thus, the
differential cross-correlations between leads 1 and 2 are negative, and in the
small tunneling limit \mbox{$\partial_\mu P_{12}
|_{\mu=0}=-\frac{2e^2}{h}\frac{z_{13}}{z_0} \frac{z_{23}}{z_0} \big|_{\lambda = 0} < 0$}.
This has the following characteristic transport signatures in addition to the
negative cross-correlations between different outer leads: (i) the sum of differential conductances is quantized,
\begin{equation}
\frac{\partial (I_1+I_2+I_3-I_4)}{\partial \mu}\bigg|_{\mu=0}=\frac{4e^2}{h},
\end{equation}
(ii) the sum of local Fano-factors at the outer leads $\sum_{i=1}^3
F_{ii}\big|_{\mu=0}\approx 1$ is quantized at zero bias and in the small tunneling limit.

To understand the transport signatures for weak tunneling, we note that the Hamiltonian~(\ref{eq:effective4}) containing the Majorana operators $\gamma_{1,2,3,4}$ can be rewritten in a
diagonalized form as $H_M=2 \xi(\psi^\dagger \psi-\tfrac{1}{2})$. The four-dimensional Hilbert space is spanned by the Dirac fermion $\psi$ and two degenerate zero-energy Majorana fermions orthogonal to it,
\begin{align}
\gamma_{A,B} =\frac{\epsilon_{14}\gamma_{3,2}-\epsilon_{(3,2)4}\gamma_1}{\sqrt{\epsilon_{14}^2+\epsilon_{(3,2)4}^2}},\quad&\label{eq:gammaB}
\psi^\dagger =\frac{1}{2\xi}\left(\sum_{i=1}^3 \varepsilon_{i4}\gamma_i+i\xi\gamma_4 \right).
\end{align}
In this notation, $\xi$ is the energy of the fermionic state $\psi$, which explains the RL at
$\mu\approx\pm2\xi$. The Dirac fermion $\psi$ is delocalized between
the three ends of the T~junction and the region near
the crossing point. The zero-energy Majorana states $\gamma_{A,B}$, on the
other hand, are delocalized among the outer MBS $\gamma_{1,2,3}$ but do not
involve the central MBS $\gamma_4$.

This explains the presence of a zero-bias differential
conductance peak at the outer leads and its absence at the central
lead. In order to better understand the zero-bias transport properties at the central
lead, we disregard the concurring transport at the outer leads for a moment by considering only the \textit{total} current $\hat{I}_o=\sum_{j=1}^3 \hat{I}_j$ transferred into the superconductor
from \textit{any} of the outer normal leads, and use the corresponding counting field $\lambda_o = \sum_{i=1}^3 \lambda_i$. The CGF~\eqref{eq:MR} then simplifies to a sum of a term describing Andreev reflection processes at outer leads, and a statistically
independent term describing non-resonant CAR~\eqref{eq:CAR} with probability
$p=(z_4/z_0)|_{\lambda=0}$ between the central lead on the one
hand, and all of the outer leads on the other hand (see Appendix~\ref{appendix:FCS_interpretation}).
This has the following characteristic transport signatures: (i) the differential
conductance at the fourth lead, $\frac{\partial I_{4}}{\partial
  \mu}|_{\mu=0}\approx 0$ is suppressed, (ii) the local Fano-factor at the
fourth lead $F_{44}\approx 1$ is quantized at zero bias and in the small
tunneling regime and (iii) a generalized non-local Fano-factor between the
central and outer leads $F_{o4}= P_{o4}/I_4|_{\mu=0} \approx 1$ is positive and quantized.\par

We have checked the validity of the effective
Hamiltonian~\eqref{eq:effective4} and of our analytical results for the
average currents and the cross-correlations by comparing them to numerical
simulations of a T~junction in a realistic semiconductor-based
setup,\cite{Reuther2013,Oreg2010,Sau2010} and we have also included weak
disorder in these simulations. The results are in excellent agreement with the
analytical predictions. They are shown in~\figref{fig:fourleadscond} and
described in detail in the Appendix~\ref{appendix:MicroscopicModel}.
We have checked that the
characteristic conductance signatures remain visible up to temperatures on the
order of $\xi$, see~\figref{fig:fourleadscond}. In addition, the different
signs of the cross-correlations and Fano-factors should be measurable in the voltage range $kT<\mu$ where shot
noise is dominant.

\section{Conclusions}
In summary, we have investigated electron transport in a generic network of Majorana bound states (MBS). We derived a general formula for the cumulant generating function for charge transport in a network of MBS, and found that it allows a clear distinction of the various transport mechanisms, i.e., crossed Andreev reflection, resonant level behavior, and local Andreev reflection. Using the general result, we predicted and interpreted the transport properties of a
T~junction of Majorana wires. We found that the hybridization of the MBS leads
to a striking distinction between the transport through the outer MBS and
transport through the central one. At small energies, while the sum of the differential
  conductances at the outer leads is quantized, transport through the central
  lead is suppressed. Moreover, current cross-correlations involving the central MBS have opposite sign from those which do not. These features
  can be physically interpreted by representing the exact eigenstates of a
  topological T~junction in terms of two nonlocal Majorana fermions and one nonlocal Dirac fermion.
We would like to point out that this effect occurs not only in T~junctions,
but generically in star-shaped networks of MBS. The hybridization of the outer
MBS with a central one gives rise to nonlocal Majorana states and has a strong
effect on the transport properties.\\
\textit{Note added}.
During the resubmission process, we became aware of a related publication,\cite{Zhou2014} which in contrast to our work
concentrates on topological phases of finite-width topological T~junctions in
the limit of long wires.

\acknowledgments
L.W. and P.R. acknowledge financial support from the NTH school for contacts in
nanosystems, the DFG grant No. RE 2978/1-1 and the EU-FP7 project SE2ND
[271554]. T.L.S. acknowledges financial support from the Swiss National Science Foundation.

\appendix

\section{FCS calculation}
\label{appendix:FCS_calculation}
Here, we present a more detailed derivation of the FCS formula in Eq.~\eqref{eq:generalFCS} in the main text. 
We can formally write the partition function [Eq.~\eqref{eq:lnchi_K} in the main text]
\begin{align}\label{lnchi_K}
 \ln \chi(\lambda) &= \ln \expct{ \TC \exp \left[ - i \int_\C ds H^{\lambda}(s) \right] }{\rho_0}
\end{align}
in the continuum notation as a functional integral\cite{Kamenev2011}
\[\chi=\int D[\hat{\gamma}] D[\bar{\hat{\psi}}\hat{\psi}] \exp{(iS[\bar{\hat{\psi}},\hat{\psi},\hat{\gamma}])}\]
The variables
$\hat{\psi}_\alpha,\bar{\hat{\psi}}_\alpha$ are mutually independent complex Grassmann
variables, while $\hat{\gamma}_\alpha$ are real fermionic Grassmann variables. The path integral for $n$ Majorana
fermions is constructed as a complex fermionic path integral by introducing an
extra set of $n$ free spectator Majoranas that are not contained in the
Hamiltonian and whose degrees of freedom are then integrated out. \par
The Keldysh action of our system is given by the lead action, the combined source-tunneling part and the Majorana action as $S[\bar{\hat{\psi}},\hat{\psi},\hat{\gamma}]=S_{\rm leads}[\hat{\psi}]+S_{\rm tun}[\hat{\psi},\hat{\gamma}]+S_{M}[\hat{\gamma}]$.
The individual components of the action are
\begin{align}
S_{M}[\hat{\gamma}]&=\sum_{\alpha,\beta} \int_C ds\,  \hat{\gamma}_\alpha(s)[\hat{D}_{0}^{-1}]_{\alpha\beta}(s)\hat{\gamma}_\beta(s)\\
S_{\rm tun}[\hat{\psi},\hat{\gamma}]&=  -i\sum_{\alpha} \int_C ds\, [t_\alpha \hat{\gamma}_\alpha(s) \bar{\hat{\psi}}_\alpha(s) e^{i\lambda_\alpha(s)/2}  - \hc ]\\
S_{\rm leads}[\hat{\psi}]&=\sum_{\alpha} \iint_C dsds' \,
\bar{\hat{\psi}}_\alpha(s) [\hat{G}_{0}^\alpha]^{-1}(s,s')
\hat{\psi}_\alpha(s') \label{eq:localGreen}
\end{align}
where
  $\hat{D}_{0}(s)$ denotes the unperturbed Majorana Green's function and where the position-integral of the lead
  action has already been performed\cite{Kamenev2011}, yielding the local
  Green's function $\hat{G}_0^\alpha(s,s')$ of lead $\alpha$. Note that the integration over the closed Keldysh contour can equivalently be
  replaced by an integration over the usual time contour by rewriting the
  respective integrand using explicit Keldysh indices on the complex fermion, Majorana and counting variables, e.g. $\hat{\psi}(s)=\hat{\psi}^\pm(s)$ for $s
  \in C_\pm$. Equivalently to this, Keldysh-rotated components
\begin{align}
\bar{\psi}_\alpha&=(\bar{\psi}_\alpha^{cl},\bar{\psi}_\alpha^q)=(\bar{\hat{\psi}}_\alpha^+-\bar{\hat{\psi}}_\alpha^-,\bar{\hat{\psi}}_\alpha^++\bar{\hat{\psi}}_\alpha^-)/\sqrt{2},\\
\psi_\alpha&=(\psi_\alpha^{cl},\psi_\alpha^q)^T=(\hat{\psi}_\alpha^++\hat{\psi}_\alpha^-,\hat{\psi}_\alpha^+-\hat{\psi}_\alpha^-)^T/\sqrt{2},\\
\gamma_\alpha&=(\gamma_\alpha^{cl},\gamma_\alpha^q)^T=(\hat{\gamma}_\alpha^++\hat{\gamma}_\alpha^-,\hat{\gamma}_\alpha^+-\hat{\gamma}_\alpha^-)^T/\sqrt{2},
\end{align}
can be used. Rewriting everything in Keldysh-rotated components, the local
Green's function of lead $\alpha$ assumes the matrix form\cite{Kamenev2011}
\[iG_0^\alpha(t,t')=\pi \rho_0 \left(\begin{matrix} \delta(t-t') & 2F_\alpha(t-t')\\ 0 & -\delta(t-t') \end{matrix}\right)\\
\] where $\rho_0 = 1/(2 \pi v_F)$ is the density of states, the Fourier transform of the distribution matrix is
$F_\alpha(\epsilon)=1-2n_\alpha(\epsilon)$ and $n_\alpha$ is the occupation number in lead
$\alpha$.
The inverse of the unperturbed Majorana Green's functions assumes the matrix form
\begin{align}
[D_{0}(t,t')^{-1}]_{\alpha\beta}&= \frac{1}{2}\delta(t-t')\left[i\partial_t\delta_{\alpha\beta}+2i\epsilon_{\alpha\beta}\right]\sigma_x
\end{align}
in Keldysh-rotated basis, where $\sigma_x$ is the first Pauli matrix in $(cl,q)$ Keldysh space and we have neglected a small regularization that would be necessary to account for the
boundary conditions of free Majorana fields.\par
The source-tunneling part of the action becomes in rotated Keldysh space
\begin{align}
  iS_{\rm tun}[\psi,\gamma]=  \sum_{\alpha} \int dt\,
  \gamma_\alpha(t)^T  \left[ \Lambda_\alpha(t)
    \bar{\psi}_\alpha^T(t)   + \sigma_x\Lambda^\dagger_\alpha(t)\psi_\alpha(t)\right]\notag
\end{align}
with
$\Lambda_\alpha(t)=t_\alpha e^{i\lambda_\alpha^{cl}(t)}e^{i\sigma_x\lambda_\alpha^q(t)}$, $(\lambda_\alpha^{cl},\lambda_\alpha^q)^T=(\lambda_\alpha^++\lambda_\alpha^-,\lambda_\alpha^+-\lambda_\alpha^-)^T$,
and where $\sigma_x$ is the first Pauli matrix.
Integrating out the degrees of freedom of the leads using a Gaussian integral for complex Grassmann variables of the form
\[\int d\eta^\dagger \int d\eta e^{-\eta^\dagger A\eta-\eta^\dagger J-J^\dagger\eta}=\det(A)e^{J^\dagger A^{-1} J}\]
reduces the problem to the Majoranas only, yielding a dissipative action
\begin{align}
S_{\rm diss}[\gamma]=\sum_{\alpha\beta} \iint dt dt' \,   \gamma^T_\alpha(t) [D^\lambda]^{-1}_{\alpha\beta} (t,t')\gamma_\beta(t'),
\end{align}
where $[D^\lambda]^{-1}=D_{0}^{-1}-\Sigma^\lambda$
and the generalized self-energy in the leads is given in rotated Keldysh basis by
\begin{align}
\Sigma_{\alpha\beta}^{\lambda}(t,t')=\delta_{\alpha\beta}
 \left[\sigma_x\Lambda_\alpha^\dagger(t)  G_0^\alpha(t,t') \Lambda_\alpha(t')\right]
\end{align}
from which after symmetrization and keeping only the quantum parts of the counting fields $\lambda_\alpha^q$, we obtain
\begin{align}
\Sigma^\lambda_{\alpha\beta}&=-i\delta_{\alpha\beta}\Gamma_\alpha
\left(\begin{matrix} \Sigma_\alpha^{qq} & \Sigma_\alpha^{q\,cl}\\
-\Sigma_\alpha^{q\,cl}& \Sigma_\alpha^{cl\,cl}\end{matrix}\right)\\
\Sigma_\alpha^{qq}(t,t')&=F^\alpha_-(t,t')(\cos\lambda^q_{\alpha,-}-\cos\lambda^q_{\alpha,+})/2\\
\Sigma_\alpha^{q\,cl}(t,t')&=\delta(t,t')\cos\lambda^q_{\alpha,+}\notag\\&+i[F^\alpha_+(t,t')\sin\lambda^q_{\alpha,+}+F^\alpha_-(t,t')\sin\lambda^q_{\alpha,-}]/2\\
\Sigma_\alpha^{cl\,cl}(t,t')&=F^\alpha_-(t,t')(\cos\lambda^q_{\alpha,-}+\cos\lambda^q_{\alpha,+})/2
\end{align}
where we have have defined $\Gamma_\alpha=2\pi\rho_0|t_\alpha|^2$,
$F^\alpha_{\pm}(t,t')=F_\alpha(t-t')\pm F_\alpha(t'-t)$ and
$\lambda^q_{\alpha,\pm}(t,t')=\lambda^q_\alpha(t)\pm\lambda^q_\alpha(t')$.

Note that the Majorana Green's function $D^\lambda$ is skew-symmetric,
\[D^\lambda_{\alpha\beta}(t,t')=-[D^\lambda_{\beta\alpha}(t',t)]^T\]
where the transpose is taken in Keldysh-space and we have allowed for integration by parts.

Integrating out the Majorana Grassmann variables using a Gaussian integral for real Grassmann variables and a skew-symmetric matrix A of the form
\[\int d^n \theta
e^{-\frac{1}{2}\theta_iA^{ij}\theta_j}=\sqrt{\det{A}}\]
yields $\chi=\sqrt{\det([D^{\lambda}]^{-1})}/\sqrt{\det([D^{\lambda=0}]^{-1})}$
for the generating function,
where the determinant is to be understood over
the (q,cl)-Keldysh indices, the lead indices, as well time, and we have enforced the
normalization $\chi(\lambda=0)=1$ explicitly. \\
We now set $\lambda^q_{\alpha}(t)=\lambda_{\alpha}\theta(t)\theta(\T-t)$, so
that for $t \in (0,\T)$, $[D^\lambda]^{-1}$ only depends on time differences. \par
For a long measurement time $\T$, the leading contribution to the cumulant
generating function is given by
\[\ln \chi(\lambda)=\frac{1}{2} \sum_\omega \ln \frac{\det [D^\lambda(\omega)]^{-1}}{\det [D^{\lambda=0}(\omega)]^{-1}}\]
where the determinant is to be understood over
the (q,cl)-Keldysh indices and the lead indices.
The energy is quantized in
units of $2\pi/\T$, where $\T$ is the measurement time.\par
The components of the self-energy in energy space are given by
\begin{align}
\Sigma_\alpha^{qq}(\omega)&=[n^\alpha(\omega)-n^\alpha(-\omega)][1-\cos(\lambda_\alpha)]\\
\Sigma_\alpha^{q\,cl}(\omega)&=e^{i\lambda_\alpha}-i[n^\alpha(\omega)+n^\alpha(-\omega)]\sin\lambda_\alpha\\
\Sigma_\alpha^{cl\,cl}(\omega)&=[n^\alpha(\omega)-n^\alpha(-\omega)][1+\cos(\lambda_\alpha)].
\end{align}
In the case of positive symmetric bias $\mu_\alpha=\mu$ and zero temperature, only energies $-\mu<\omega<\mu$ contribute to the cumulant generating function. Performing the determinant over
Keldysh indices, we can further simplify this to
\[\ln \chi(\lambda)= \T\int_{-\mu}^\mu \frac{d\omega}{2\pi}\ln
\det[1-G_R(\omega)V^\lambda_R(\omega)]\]
where the determinant is to be understood only over lead indices $\alpha,\beta$. The counting-field-independent component $G_R^{-1}$ and the counting-field-dependent component $V_R^\lambda$ of the retarded inverse Majorana Green's function at zero temperature, are given by, respectively
\begin{align}
[G_R(\omega)^{-1}]_{\alpha\beta}&=\delta_{\alpha\beta}\left(\frac{\omega}{2}+i\Gamma_\alpha\right)+i\epsilon_{\alpha\beta}\\
[V^\lambda_R(\omega)]_{\alpha\beta}&=-i\delta_{\alpha\beta}\Gamma_\alpha (e^{-i\lambda_\alpha}
-1).
\end{align}

\section{Interpretation of the FCS at zero bias}

\label{appendix:FCS_interpretation}
Here, we show that the CGF Eq.~\eqref{eq:MR} in the main text
can be written as a sum of two statistically independent processes if we consider only the current $\hat{I}_4$ at lead 4 and the \textit{total} current at the outer leads, $\hat{I}_o=\sum_{j=1}^3 \hat{I}_j$: two overall Andreev reflection (AR) processes at the
outer leads and a crossed Andreev reflection (CAR) process between the outer leads and the central lead. In
order to see this, we start from the formula for the CGF in the CAR limit $\mu \to 0$ at
finite $\epsilon_{\alpha\beta}$  [see Eq.~\eqref{eq:MR} in the main text],
\begin{align}
\frac{\partial}{\partial \mu} \ln \chi\big|_{\mu= 0} &=
\frac{\T}{\pi} \ln \left[ \frac{z_{12} + z_{23} + z_{31} + z_4}{z_0} \right], \label{eq:MRS}
\end{align}
where $z_0 = (z_{12} + z_{23} + z_{31} + z_4)|_{\lambda = 0}$, $z_{12} = \Gamma_1 \Gamma_2 \epsilon_{34}^2 e^{-i(\lambda_1+\lambda_2)}$, and
$z_{23}$, $z_{31}$ are obtained by cyclic permutations of the indices $1,2,3$ in the
formula for $z_{12}$. In addition, $z_4 =\Gamma_1\Gamma_2\Gamma_3\Gamma_4e^{-i(\lambda_1+\lambda_2+\lambda_3+\lambda_4)}$. Considering only the \textit{total} current $\hat{I}_o$ transferred into the SC
from \textit{any} of the outer normal leads, we introduce the counting field $\lambda_o = \lambda_1 + \lambda_2 + \lambda_3$. Thereby, the CGF~\eqref{eq:MRS} is simplified to
\begin{align}
& \frac{\partial}{\partial \mu} \ln \chi\big|_{\mu= 0} =
 \frac{\T}{\pi} \ln \left[ \frac{z_{12} + z_{23} + z_{31} + z_4}{z_0} \right] \notag \\
&=\frac{\T}{\pi} \ln \left[ \frac{ \left(z_{12} + z_{23} + z_{31}
    \right)|_{\lambda =   0} e^{-2i\lambda_o}  + z_4|_{\lambda =
      0}e^{-i\lambda_4-3i\lambda_o}}{z_0} \right] \notag \\
&=\frac{\T}{\pi}\left\{-2i\lambda_o +\ln \left[\frac{ z_0   + z_4|_{\lambda =
      0}(e^{-i\lambda_4-i\lambda_o}-1)}{z_0}\right]\right\}\notag \\
&=\frac{\T}{\pi}\left\{-2i\lambda_o +\ln \left[1   + p(e^{-i\lambda_4-i\lambda_o}-1)\right]\right\}
\end{align}
where $p = z_4|_{\lambda = 0}/z_0$. The first term describes two independent AR processes at the outer leads,
leading to the differential conductance
\begin{align}
  \partial_\mu
I_o=\frac{2e^2}{h} (2 + p) \approx \frac{4 e^2}{h}
\end{align}
for the total conductance at the three outer
leads.
In contrast, the second term describes a CAR process between the outer and the
central lead with probability $p$. Interestingly, this
process only takes place when $\Gamma_\alpha\neq 0$ for all $\alpha=1,2,3,4$, and it is suppressed
by a factor $\Gamma_\alpha/\epsilon_{\beta\beta'}$ for small tunneling. Therefore, the differential conductance at the central lead is
small,
\begin{align}
  \partial_\mu I_4=\frac{2e^2}{h}p\approx 0
\end{align}
The local Fano factor at zero bias is given by
\[F_{44}= \frac{p(1-p)}{p}\approx 1\]
and the generalized nonlocal Fano factor between the
central and outer leads is given by
\[F_{o4}= \frac{P_{o4}}{I_4}\bigg|_{\mu=0} =\frac{p(1-p)}{p}\approx 1.\]

\section{Physical realization}
Here, we present the low-energy solutions of two microscopic physical models for the T~junction to explicitly determine the MBS at the outer ends and at the crossing point: we first solve analytically the Kitaev chain model
generalized to two crossed 1D wires, and then present a numerical solution using a discretized
version of the full Hamiltonian of a T~junction with finite
width in a semiconductor nanowire. This allows us to validate the effective Hamiltonian Eq.~\eqref{eq:effective4} in the
main text.

Moreover, we use the microscopic Hamiltonian to determine the transport properties numerically and show that the numerical results fit extremely well with the result from the FCS
determinant formula~\eqref{eq:FCSzeroTemp} applied to the low-energy
Hamiltonian~\eqref{eq:effective4} as presented in the main text, even in the
presence of weak disorder. These results strongly suggest that the transport
properties of topological T~junctions presented in the main text should be
measurable in real devices.

\label{appendix:physical_realization}
The setup described in the main text could be realized by
patterning of HgTe quantum wells\cite{Reuther2013} or by crossed semiconductor quantum wires.
Consider for example a 2DEG in the $xy$-plane subject to the pair potential $\Delta$ of an $s$-wave superconductor and a Zeeman-field $h$ in $z$-direction, as well as Rashba
spin-orbit coupling $u$. The Hamiltonian of this two-dimensional system is given in the discrete notation by
\begin{align}
H=\frac{1}{2}
\sum_{i,j} \left[\Psi^\dagger_{ij} H^0 \Psi_{ij} + \left(\Psi^\dagger_{ij} t_x \Psi_{i+1,j} +\Psi^\dagger_{ij}  t_y \Psi_{i,j+1}+ \hc\right)\right]
\label{eq:H2DEGdisc}
\end{align}
where the onsite Hamiltonian $H^0$ and the hopping matrices $t_x, t_y$ are given by
\begin{align}
H^0&=\left(  -\mu+\frac{1}{ma^2} \right)\tau_z+h\sigma_z+ \Delta \tau_x \\
t_x&=-\frac{1}{2ma^2}\tau_z-\frac{i}{2a}u \sigma_y\tau_z\\
t_y&=-\frac{1}{2ma^2}\tau_z+\frac{i}{2a}u \sigma_x\tau_z
\end{align}
and $a$ is the lattice constant and $\sigma_{x,y,z}$ and $\tau_{x,y,z}$ are Pauli matrices in spin and Nambu space, respectively. For simplicity we assumed that $\Delta$ is real.

In the following, we seek low-energy solutions for the case when the 2DEG is patterned into quantum wires in the shape of a T, with one wire
along the $x$-axis, the other along the $y$-axis (setup shown in Fig.~\ref{fig:DOSkwant}). These solutions can be calculated analytically in the limit where the system can be mapped onto two crossed Kitaev chains, or numerically for more general parameters.

\subsection{Kitaev chain}
\label{appendix:Kitaev}
We first present an analytical solution of~\eqref{eq:H2DEGdisc} in the small-width and ``effective $p$-wave'' limit, where we can map the problem onto two crossed Kitaev chains with effective parameters $\Delta^{\eff}, t^{\eff}, \mu^{\eff}$.\cite{Alicea2010} As the chains intersect at an angle of $\pi/2$, their superconducting phases differ by $\pi/2$, as can be calculated explicitly by following the procedure in Ref.~[\onlinecite{Alicea2010}]. This yields the following Kitaev model for the topological T~junction:

\begin{align}
H= &\sum_{j=-N_L}^{N_R} \left[ -\mu^\eff \left(c_{j,0}^\dagger   c_{j,0}-\frac{1}{2}\right)\right]\notag\\
&+\sum_{j=-N_L}^{N_R-1}\left[\left(-\frac{t^\eff}{2} c_{j,0}^\dagger
    c_{j+1,0}+\frac{\Delta^\eff}{2} c_{j,0}
    c_{j+1,0}+\hc\right)\right]\notag\\
+&\sum_{k=1}^{N_D}\left[ -\mu^\eff \left(c_{0,k}^\dagger   c_{0,k}-\frac{1}{2}\right)\right]\notag\\
&+ \sum_{k=0}^{N_D-1}\left[\left(-\frac{t^\eff}{2} c_{0,k}^\dagger c_{0,k+1}+\frac{-i\Delta^\eff }{2} c_{0,k} c_{0,k+1}+\hc\right)\right]
\end{align}
where $c_{i,j}^\dagger$ creates an electron at position $i,j$ and $N_L,N_R,N_D$ denote the numbers of sites of the left, right, and vertical section, respectively, of the T~junction.
At the ``sweet spot'' $\Delta^{\eff}=t^{\eff}$ and $\mu^{\eff}=0$, the Hamiltonian has a bulk gap of $t^{\eff}/2$ and
four MBS $\tilde{\gamma}_\alpha$ at zero energy, three of which lie at
the edges of the T~junction and one lies at the crossing point,
\begin{align}
\tilde{\gamma}_1&=c_{-N_L,0}+c_{-N_L,0}^\dagger\\
\tilde{\gamma}_2&=-i(c_{N_R,0}-c_{N_R,0}^\dagger)\\
\tilde{\gamma}_3&=e^{-i\pi/4}c_{0,N_D}+e^{i\pi/4}c_{0,N_D}^\dagger\\
\tilde{\gamma}_4&=-\frac{i}{2}(c_{-1,0}-c_{-1,0}^\dagger)\notag-\frac{1}{2}(c_{1,0}+c_{1,0}^\dagger)\\&+\frac{i}{\sqrt{2}}(e^{-i\pi/4}c_{0,1}-e^{i\pi/4}c_{0,1}^\dagger)\label{eq:gamma4}
\end{align}
Away from the ``sweet spot'', the MBS
$\tilde{\gamma}_\alpha$ are only exponentially localized, resulting in a
hybridization that decreases exponentially with the chain
length.\cite{Kitaev2001} This results in the characteristic spectrum of a
nonlocal Dirac fermion at nonzero energy $\xi$ and two degenerate Majorana
levels at energy close to zero, as described by the effective Hamiltonian Eq.~\eqref{eq:effective4} in the main text.

In the special case where three arms of the T~junction are of equal length $N=N_L=N_R=N_D$,
the low-energy Hamiltonian describing the four Majoranas of the topological T~junction is given by
\begin{equation}
H_M=i\left(\frac{\epsilon}{2}\gamma_1\gamma_4+\frac{\epsilon}{2}\gamma_2\gamma_4+\frac{\epsilon}{\sqrt{2}}\gamma_3\gamma_4\right). \label{eq:eff}
\end{equation}
with energy splitting $\epsilon$, that is proportional to the overlap of
adjacent MBS. The coefficients of $1/2$ and $1/\sqrt{2}$ are a result of the specific form of the MBS $\gamma_4$ at the junction, whose wavefunction decays exponentially into all three wires.

\subsection{Microscopic model}
\label{appendix:MicroscopicModel}
\paragraph{Validity of effective Hamiltonian.} First, we discuss the validity
of the effective Hamiltonian~\eqref{eq:eff} in a microscopic model~\eqref{eq:H2DEGdisc}. We choose the parameters as follows: Throughout the entire setup, we use a Zeeman energy of $h=1.5$\,meV (corresponding to a g-factor of $g=50$ and a magnetic field of $B=1$\,T), a Rashba spin-orbit coupling $u=50$\,meV\,nm and an effective mass of $m=0.015m_e\approx 1.6 \cdot 10^{-4}\,\hbar^2/(\mathrm{nm}^2\,\mathrm{meV})$. For the T~junction region, we use an induced proximity gap $\Delta=0.5$\,meV, and a chemical potential of $\mu=0$.

These parameters compare well with existing experiments.\cite{Mourik2012,Churchill2013} We have checked that the results that will be presented in the following do not depend on the choice of the lattice constant $a$ for $a<0.01\,\mum$.

The topological properties of a thin wire of the
Hamiltonian~\eqref{eq:H2DEGdisc} strongly depend on its width. As the width
increases, it undergoes a series of topological transitions as shown
in Refs.~[\onlinecite{Kells2012, Rieder2012}] in the ``effective $p$-wave'' limit. We choose
the width as $w = 0.1\,\mum$, which is of the same order as the diameter of
spin-orbit coupled quantum wires used in existing
experiments,\cite{Mourik2012,Churchill2013} and for which we have checked that a long wire with hard-wall boundary conditions has MBS at its ends. The splitting energy of a Majorana wire depends on the parameters and decays in an oscillating and exponentially decaying way with the length of the wire.\cite{Rainis2013,DasSarma2012} We choose the length of the wire as $l = 1.25\,\mum$, so that we are away from the minimum of the oscillation. \par
We find that for these parameters, the splitting of two Majoranas in a Majorana wire of length $l = 1.25\,\mum$ is $\epsilon \approx 8\mueV$, while the splitting of a Majorana wire of length $2l = 2.5\,\mum$ is $\epsilon \ll 1\mueV$.

\begin{figure}[ht]
    \includegraphics[width=0.98\columnwidth]{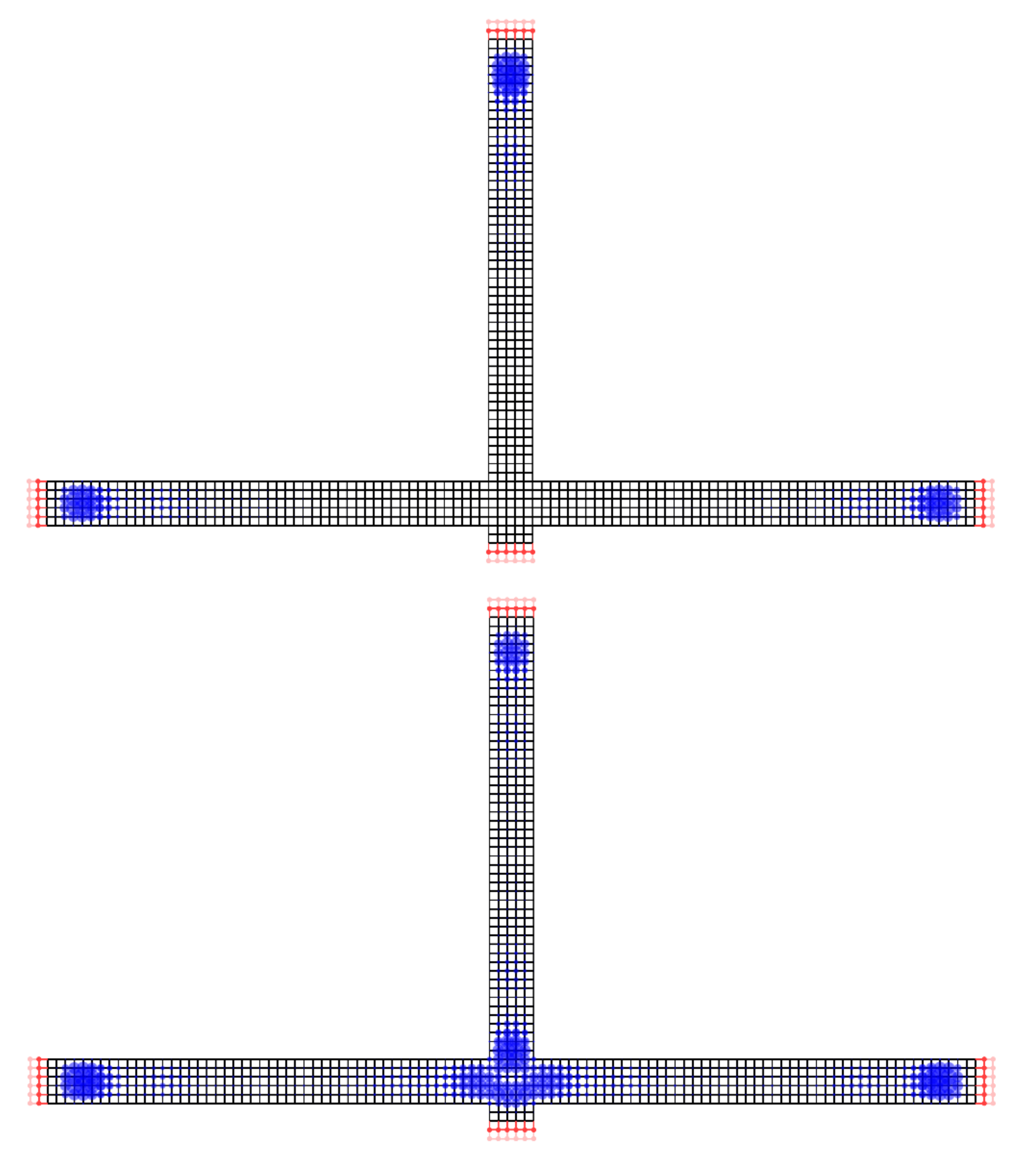}
    \centering
    \caption{
Electronic probability density in a Rashba 2DEG-based effective topological
superconductor patterned into a T~junction. The size of the blue point around
the lattice sites shows the mean probability density on a given lattice site. \emph{Top graph:} Majorana states at energy $E \approx 0$. \emph{Bottom graph:} Nonlocal Dirac fermion at nonzero energy $E \approx \xi$.}
    \label{fig:DOSkwant}
\end{figure}

Subsequently, we pattern the 2DEG in the form of a T~junction with three arms
of equal length $l = 1.25\mum$ and calculate the lowest energy wavefunctions
and the transport properties. Indeed, the low-energy spectrum and
wave-functions are qualitatively similar to the prediction from two crossed
Kitaev wires in the previous section. As shown in Fig.~\ref{fig:DOSkwant}, there
is a nonlocal Dirac fermion at nonzero energy $E \approx \xi$ located near
the crossing point and at the ends of the wires, as well as MBS at
energy $E \approx 0$ and localized only at the ends of the wires. We estimate
the energy of the nonlocal fermion of the T~junction from the numerical data
as $\xi=8\,\mueV$. The numerical data is indeed consistent with a splitting of
adjacent Majoranas of $\epsilon_{14}=\epsilon_{24}=\frac{\xi}{2},\epsilon_{34}=\frac{\xi}{\sqrt{2}}$,
as in the calculation for the crossed Kitaev wire~\eqref{eq:eff}, and a vanishing splitting
energy for non-nearest-neighbor MBS, thereby validating the effective
Hamiltonian in Eq.~\eqref{eq:effective4} in the main text.

\paragraph{Transport calculation.} In the following, we study the transport properties in a realistic T~junction setup.
In the transport setup, each wire end in the T~junction is connected by a
tunnel barrier of length $l_B = 0.1\,\mum$ and height $\mu=-20\,\mueV$ to a
lead with periodic boundary conditions in lateral direction and chemical
potential $\mu=0$. Due to this symmetry for the outer leads, their effective
tunnel couplings $\Gamma_1=\Gamma_2=\Gamma_3=\Gamma$ are all identical. We
extract this tunnel coupling as $\Gamma=1.62\,\mueV$ by considering a single
lead coupled to a simple Majorana wire. Finally, we determine the tunnel
coupling to the Majorana at the crossing point from the numerics when only the
fourth lead is attached to the T~junction to $\Gamma_4\approx\Gamma/4$.

For the above choice of parameters, the splitting between adjacent Majoranas
is larger than the tunnel coupling and substantially larger than the
next-to-nearest neighbor coupling, so that we really are in the regime
where Eq.~\eqref{eq:effective4} in the main text is a good approximation. We
now proceed to directly compare the predictions of the FCS calculation based
on the FCS determinant formula Eq.~\eqref{eq:FCSzeroTemp} applied to the low-energy Hamiltonian~\eqref{eq:effective4} as presented in the main text with the transport properties of the microscopic model. The latter can be calculated numerically using the open source code KWANT.\cite{Groth2013} Figure~\ref{fig:fourleadscond} shows that the FCS calculation and the calculation for the microscopic model are in excellent agreement.

\paragraph{Weak Gaussian disorder.} We also study the effect of on-site disorder on the transport properties by including weak (compared to the bulk gap) Gaussian fluctuations of the chemical potential with standard deviation $\sigma=10\,\mueV$. We then calculate the transport properties numerically for each disorder realization and perform the ensemble average. Figure~\ref{fig:fourleadscond} shows that the results of the microscopic model remain stable also in the presence of weak Gaussian disorder in the local chemical potential.

\bibliographystyle{apsrev}

\end{document}